\documentclass[conference]{IEEEtran}
\IEEEoverridecommandlockouts
\usepackage{amsmath,amssymb,amsfonts,amsthm}
\usepackage{siunitx}
\usepackage{algpseudocode,algorithm,algorithmicx}
\usepackage{graphicx}
\usepackage{epstopdf}
\usepackage[pdf]{pstricks}
\usepackage{svg}
\usepackage{bm}
\usepackage{textcomp}
\usepackage{tikz}
\usepackage{xcolor}
\usepackage{enumitem} 
\usetikzlibrary{plotmarks}
\usepackage{pgfplots}
\usepackage{caption}
\usepackage{subcaption}
\usetikzlibrary{calc,fit,arrows,plotmarks,intersections,patterns,shapes,decorations,positioning, backgrounds, matrix}
\usepackage{lipsum}
\usetikzlibrary{arrows.meta}
\usepackage{mwe}
\usepackage{xcolor}
\usepackage{cite}
\makeatletter
\def\blfootnote{\xdef\@thefnmark{}\@footnotetext}
\makeatother
\def\BibTeX{{\rm B\kern-.05em{\sc i\kern-.025em b}\kern-.08em
		T\kern-.1667em\lower.7ex\hbox{E}\kern-.125emX}}
\usepackage{amsthm}
\usepackage{mathtools}
    \title{Short Blocks, Fast Sensing: Finite Blocklength Tradeoffs in RIS-Assisted ISAC\\
    \thanks{This work was supported by the German Research Foundation (“Deutsche Forschungsgemeinschaft”) (DFG) under Project–ID 449601577 and under Project–ID 287022738 TRR 196 for Project S03. \\$^{\star}$Authors contributed equally.}
}
	\author{\IEEEauthorblockA{Adam Umra$^{\dagger}$$^{\star}$, Kevin Weinberger$^{\dagger}$$^{\star}$, Aymen Khaleel$^{\dagger}$, Gerald Enzner$^{\ddagger}$, and Aydin Sezgin$^{\dagger}$}
		\IEEEauthorblockA{
			$^{\dagger}$Ruhr University Bochum, Germany \\
            $^{\ddagger}$Carl von Ossietzky University of Oldenburg, Germany \\
			Email: {\{adam.umra, kevin.weinberger, aymen.khaleel, aydin.sezgin\}@rub.de}, gerald.enzner@uol.de}
                
	}

\begin{document}
	\maketitle
 %================= Abstract =================% 
	\begin{abstract}
   	    Integrated sensing and communication (ISAC) is a cornerstone for future sixth-generation (6G) networks, enabling simultaneous connectivity and environmental awareness. However, practical realization faces significant challenges, including residual self-interference (SI) in full-duplex (FD) systems and performance degradation of short-packet transmissions under finite blocklength (FBL) constraints. This work studies a reconfigurable intelligent surface (RIS)-assisted FD ISAC system serving multiple downlink users while tracking a moving target, explicitly accounting for SI and FBL effects in both communication and sensing. We formulate an optimization framework to minimize service adaptation gaps while ensuring sensing reliability, solved via alternating optimization and successive convex approximation. Numerical results show that short blocklengths enable fast adaptation but increase radar outage due to fewer pulses and motion sensitivity, whereas longer blocklengths improve SINR and reduce outages but allow target motion to degrade sensing. A regime-dependent “sweet spot” emerges, where blocklength and beamformer allocation jointly balance throughput and sensing, with HD favored at very short blocklengths and FD at longer blocklengths. RIS-assisted optimization expands the FD-operating regime, enabling reliable communication and radar sensing at lower blocklengths.
	\end{abstract}

  %================= Introduction =================%  
	\section{Introduction}

The next generation of cellular networks is expected to provide ubiquitous connectivity while simultaneously enabling new functionalities such as high-precision localization, environment mapping, and object tracking~\cite{trevlakis2023localization,umra2025RL}. To meet these requirements under severe spectrum scarcity, the paradigm of \emph{Integrated Sensing and Communication} (ISAC) has recently emerged as a key enabler for sixth-generation (6G) systems \cite{liu2022ISAC,zhang20196G}. By allowing communication and sensing to share spectral, hardware, and signal-processing resources, ISAC enables a more efficient use of limited spectrum compared to the conventional design of separate systems. Beyond efficiency, ISAC also unlocks new applications in cellular networks, including real-time traffic monitoring, autonomous driving, and immersive extended reality services, where network infrastructure must support both reliable connectivity and environment awareness in a tightly integrated manner.

Despite these promising prospects, realizing practical ISAC in cellular networks faces several architectural and signal-processing challenges. Conventional approaches often rely on half-duplex (HD) operation, where sensing and communication are performed in alternating time slots or frequency bands. While this simplifies transceiver design, it inherently reduces efficiency and responsiveness, since the base station cannot sense and communicate simultaneously. A more attractive solution is to adopt a \emph{full-duplex (FD)} transceiver architecture, where the base station continuously transmits a compound waveform for downlink communication and sensing, while at the same time receiving echoes from targets in the environment \cite{xiao2021FD}. This FD operation enables uninterrupted sensing and communication, which is particularly valuable in dynamic scenarios with moving targets. However, the major impediment in such systems is the presence of \emph{self-interference} (SI) from the base station’s own transmission, which may overwhelm the received sensing echoes. Although advanced \emph{self-interference cancellation} (SIC) methods can suppress the dominant components of this interference, practical implementations always leave a non-negligible residual term~\cite{nguyen2021SI}. Understanding the impact of such residual SI on the joint performance of sensing and communication is thus essential.

In addition to transceiver impairments, a fundamental limitation arises from the assumptions underpinning classical communication theory. In particular, \emph{finite blocklength} (FBL) communication fundamentally alters the achievable rate--reliability trade-off. Classical information-theoretic results assume infinitely long codewords, which lead to sharp asymptotic capacity limits. In practice, however, many emerging services such as industrial automation, vehicular communication, and augmented reality require the transmission of short packets under strict reliability and latency constraints~\cite{polyanskiy2010FBL,durisi2016ShortPack}. In these regimes, the performance deviates significantly from the asymptotic capacity, and error probabilities are heavily influenced by blocklength~\cite{weinberger2025symbolcountsresilientwireless}. Importantly, in a FD ISAC setting the blocklength does not only affect decoding reliability; it also effectively determines the sensing dwell time (or observation interval), and thereby the number of collected echo samples and the sensitivity to target motion within that interval. Consequently, the joint presence of residual SI and FBL effects raises new questions about how communication reliability and sensing accuracy trade off in a FD ISAC system, and whether conventional design rules remain valid under such coupled latency--sensing constraints.

To further enhance coverage and controllability of the propagation environment in such tightly constrained regimes, \emph{Reconfigurable Intelligent Surfaces} (RISs) offer an appealing mechanism for shaping signal propagation via nearly passive, programmable reflection elements~\cite{basar2019wireless,weinberger2023ris}. RIS assistance is particularly relevant for FD ISAC, where improving the effective sensing and communication links can mitigate channel impairments and partially offset the losses caused by residual SI and short-packet operation \cite{li2025lightweightframeworkintegratedsensing}.

In this paper, we study a RIS-assisted FD ISAC base station serving multiple downlink users while tracking a moving target. We consider two practically relevant non-idealities, namely finite-blocklength transmission and residual self-interference, and optimize downlink performance subject to sensing-reliability constraints. The resulting non-convex problem is addressed via alternating optimization and successive convex approximation. In contrast to prior work that often assumes infinite blocklength or near-perfect SI suppression~\cite{sivadevuni2023JCAS}, our framework captures the coupled throughput--sensing tradeoff induced by latency-driven short packets.

Our main contributions are threefold. First, we formulate a joint RIS-assisted FD ISAC model that captures FBL communication, residual SI, and target motion over the sensing interval within a unified framework. Second, we introduce a sensing--communication coupling constraint that links radar SINR to communication blocklength through the sensing observation time, thereby exposing how short packets affect sensing robustness in dynamic scenarios. Third, numerical results reveal a regime-dependent blocklength \emph{sweet spot} and clarify, through FD/HD benchmarks and RIS sensitivity analyses, when RIS assistance expands the operating regime in which FD ISAC is beneficial.

%================= System Model & Problem Formualation =================%  
\section{System Model} \label{Sys_model}

\begin{figure}[t]
  \centering
  \centerline{\includegraphics[width=1\linewidth]{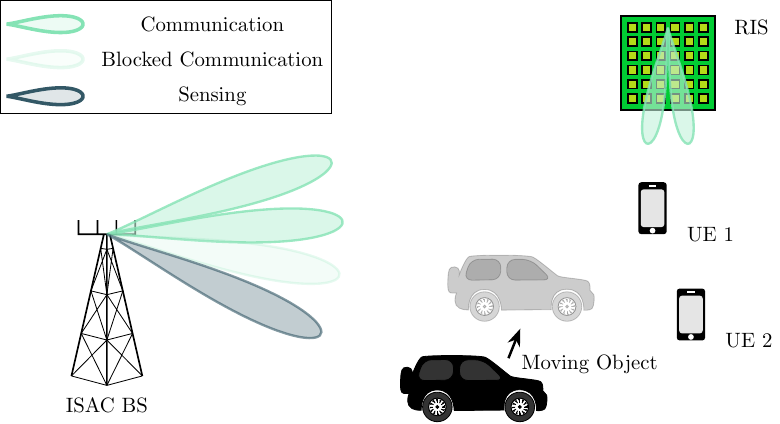}}
  % \centerline{\includegraphics[width=\linewidth]{Figures/sysmod.jpg}}
%  \vspace{2.0cm}
  \caption{RIS-assisted multiuser ISAC system with a moving object.}\label{fig:system}
\end{figure}

The base station (BS) is equipped with $N_T$ transmit antennas and $N_R$ receive antennas. 
The considered ISAC system also involves a set of single-antenna users, denoted by 
$\mathcal{K}=\{1,2,\dots, K\}$, which are served by the BS. 
Moreover, the environment contains a moving point target $s$ and a RIS consisting of $M$ reflecting elements. The system model is shown in Figure~\ref{fig:system}.
\subsection{Communication Model}
At the BS side, the composite transmit signal $\mathbf{x}$ can be expressed as
\begin{equation} \label{trns_eqn}
\mathbf{x} = \sum_{k \in \mathcal{K}} \mathbf{w}_{k} s_k + \mathbf{w}_s s_s \text{,}
\end{equation}
where $s_k, s_s \in \mathbb{C}$ denote the communication and sensing symbols, respectively, each normalized to unit average power, i.e., $\mathbb{E}[|s_k|^2] = \mathbb{E}[|s_s|^2] = 1$. These signals are transmitted simultaneously through their associated beamforming vectors 
$\mathbf{w}_k, \mathbf{w}_s \in \mathbb{C}^{N_T \times 1}$. Without loss of generality, we assume that the communication and sensing signals are uncorrelated, which can be achieved by generating the sensing signal using pseudo-random coding \cite{liu2020joint}. Accordingly, the transmit signal in \eqref{trns_eqn} establishes the basis for the joint communication and sensing operation. In the following, we specify the channel models that govern the propagation of these signals between the BS, the RIS, the users, and the target.

The considered wireless channels follow a quasi-static block fading model, where each channel remains constant within a coherence interval and changes independently across different intervals according to identically distributed random and independent processes.  

The received signal at user $k$ can be expressed as
\begin{equation} \label{Rcvd_sig}
    y_k = \left[\mathbf{h}^H_{k} + \mathbf{g}^H_{k} \operatorname{diag}(\mathbf{v^*}) \mathbf{H}\right]\mathbf{x} + n_k,
\end{equation}
where $\mathbf{h}_{k} \in \mathbb{C}^{N_T \times 1}$ denotes the baseband equivalent channel vector from the BS to user $k$, representing the direct link. The matrix $\mathbf{H} \in \mathbb{C}^{M \times N_T}$ corresponds to the BS-to-RIS channel, while $\mathbf{g}_{k} \in \mathbb{C}^{M \times 1}$ models the RIS-to-user $k$ channel. The additive white Gaussian noise (AWGN) at user $k$ is represented by $n_k \sim \mathcal{CN}(0,\sigma^2_k)$, where $\sigma^2_k$ is the noise power. The RIS-induced phase-shift matrix is given by $\operatorname{diag}(\mathbf{v}^*)$, where $\mathbf{v} = [v_1, v_2, \dots, v_M]^T$ with $v_m = e^{j\varphi_m}$, and $\varphi_m \in [0,2\pi)$ denotes the phase shift applied by the $m$-th RIS element. For notational convenience, we define the effective cascaded channel for user $k$ as
$\widehat{\mathbf{h}}_k^H \triangleq  \mathbf{h}_k^H + \mathbf{g}_k^H \operatorname{diag}(\mathbf{v}^*) \mathbf{H}$.

Based on this definition, the received SINR at user $k$ is given by
\begin{equation} \label{comm_SINR}
    \Gamma_{k} = \frac{|\widehat{\mathbf{h}}_k^H \mathbf{w}_k|^2}{\sum_{k' \in \mathcal{K}\setminus\{k\}} |\widehat{\mathbf{h}}_{k}^H \mathbf{w}_{k'}|^2 + |\widehat{\mathbf{h}}_{k}^H \mathbf{w}_s|^2 + \sigma_k^2}.
\end{equation}

In the FBL regime, quality of service (QoS) requirements are satisfied if the following condition holds~\cite{polyanskiy2010FBL}
\begin{equation}\label{eq:rateEq}
    r_k^{\mathrm{des}} \leq r_k \leq B \left( \log_2(1+\Gamma_k) - \Omega \sqrt{\frac{V(\Gamma_k)}{\eta}} \right),
\end{equation}
where $r_k^{\mathrm{des}}$ denotes the desired rate corresponding to the QoS requirement, and $r_k$ is the achievable rate of user $k$.
Here, $\eta$ denotes the blocklength, $\epsilon$ is the target block error rate (BLER), and $\Omega = Q^{-1}(\epsilon)\log_2(e)$ with $Q^{-1}(\cdot)$ denotes the inverse of the Gaussian $Q$-function $Q(x) = \int_x^\infty \frac{1}{\sqrt{2\pi}} \exp\!\left(-{t^2/2}\right) dt$, $V(\Gamma_k) = 1 - \frac{1}{(1+\Gamma_k)^2}$ is the channel dispersion and $B$ is the bandwidth. The above expression for channel dispersion is derived under the assumption of Gaussian signaling.

\subsection{Sensing Model}
For continuous target tracking, the BS operates in a monostatic radar mode. 
It is assumed that the BS is equipped with advanced signal processing capabilities, 
which enable reliable estimation of the target's direction-of-arrival (DoA) and angular velocity. 
These estimates can be obtained with sufficient accuracy as long as the sensing signal maintains an adequate SINR within each coherent processing block. In this work, the coherence time of the target’s channel is determined by the blocklength $\eta$, with one sensing block defined as
\begin{align}\label{eq:T_sens}
    T_\mathsf{sens}  = \frac{\eta}{B}.
\end{align}
Furthermore, we denote $t$ as the time slot index, and $\tau_t$ as the starting time of slot $t$. We assume that the coherence time of the moving target is smaller than the coherence of the communication users, i.e., $T_\mathsf{sens} \ll T_\mathsf{c}$. Accordingly, the total number of sensing slots within one coherence interval is given by
\begin{align}
    N_{\text{slot}} = \left\lceil \frac{T_\mathsf{c}}{T_{\text{sens}}} \right\rceil .
\end{align}

Thus, based on the blocklength parameter $\eta$, the system operates over $N_{\text{slot}}$ discrete time slots during one coherence period.

The received sensing signal at the BS can be expressed as
\begin{equation} \label{Sens_sig}
    \mathbf{y}_s = 
    \underbrace{\alpha \mathbf{a}_R(\theta)\mathbf{a}_T^H(\theta)\mathbf{x}}_{\text{Echo}}
    + \underbrace{\widetilde{\mathbf{H}}\mathbf{x}}_{\text{SI}}
    + \mathbf{n}_s,
\end{equation}
where $\widetilde{\mathbf{H}} \in \mathbb{C}^{N_R \times N_T}$ models the SI channel between the BS transmitter and receiver and $\mathbf{n}_s \sim \mathcal{CN}(\mathbf{0}, \sigma_s^2 \mathbf{I}_{N_R})$ denotes the AWGN. 

The first term in \eqref{Sens_sig} corresponds to the desired target echo, characterized by the complex coefficient $\alpha$, 
which encapsulates both the radar cross-section (RCS) of the target and the propagation pathloss. 
Specifically, $\alpha$ is expressed as
\begin{equation} \label{alpha}
    \alpha = \sqrt{\delta \left(\frac{d_0}{d_{xy}}\right)^{\nu_{xy}}},
\end{equation}
where $\delta$ is the reference pathloss at distance $d_0$, $d_{xy}$ denotes the distance between the BS and the target, and $\nu_{xy}$ is the pathloss exponent \cite{jing2024ISAC}. 
The vectors $\mathbf{a}_R(\theta) \in \mathbb{C}^{N_R \times 1}$ and $\mathbf{a}_T(\theta) \in \mathbb{C}^{N_T \times 1}$ represent the receive and transmit steering vectors corresponding to the target angle $\theta$. For notational simplicity, we define
$\mathbf{G} \triangleq \alpha \mathbf{a}_R(\theta)\mathbf{a}_T^H(\theta)$. Using this definition, the received signal in \eqref{Sens_sig} can be rewritten as
\begin{equation}
    \mathbf{y}_s = \mathbf{G}\mathbf{x}
    + \widetilde{\mathbf{H}} \mathbf{x}
    + \mathbf{n}_s.
\end{equation}

The radar echo SINR can be initially expressed as
\begin{equation} \label{rad_SINR_initial}
    \gamma_s = \frac{\|\mathbf{G}\mathbf{w}_s\|_2^2}{\|\widetilde{\mathbf{H}}\mathbf{W}\|_F^2 + \sigma_s^2},
\end{equation}
where $\mathbf{W} = \left[\mathbf{w}_1, \dots, \mathbf{w}_K, \mathbf{w}_s\right]$ denotes the transmit beamforming matrix, $\|\cdot\|_2$ is the Euclidean norm, and $\|\cdot\|_F$ is the Frobenius norm.  

In practice, even after applying SIC (including passive suppression, analog, and digital cancellation), a certain amount of residual SI remains due to hardware impairments and imperfect channel estimation. To capture this effect, we approximate the residual SI power as a fraction of the transmit power, i.e.,
\begin{equation} \label{residual_power}
    \|\widetilde{\mathbf{H}}\mathbf{W}\|_F^2 \approx \rho \widetilde{P}_\mathrm{T}(\mathbf{W}),
\end{equation}

where $\rho \in [0,1]$ characterizes the level of SI cancellation imperfection~\cite{xiao2022FD} and with
\begin{align}
\widetilde{P}_\mathrm{T}(\mathbf{W}) = \sum_{k \in \mathcal{K}}\!\|\mathbf{w}_k\|_2^2 + \|\mathbf{w}_s\|_2^2
\end{align} 
Accordingly, the effective radar SINR can be reformulated as
\begin{equation} \label{rad_SINR_final}
    \gamma_s = \frac{\|\mathbf{G}\mathbf{w}_s\|_2^2}{\rho \widetilde{P}_\mathrm{T}(\mathbf{W}) + \sigma_s^2}.
\end{equation}

    %\begin{figure}
    %    \begin{center}     \includegraphics[width=\columnwidth, height=5cm]{ResMetric.pdf}
   %     \end{center}
   %     \caption{The interplay between time and system functionality.}
    %    \label{ResMetric}
    %\end{figure}    
%================= Proposed Schemes =================%  
\section{Problem Formulation} \label{Opt} 

We now formulate an optimization problem that captures the impact of FBL transmission and residual SI on the ISAC QoS. Following the adaptation-gap metric in~\cite{reifert2022comeback,weinberger2025symbolcountsresilientwireless}, the objective minimizes the deviation between the achieved rates and the desired service targets. Moreover, similar to RIS-assisted ISAC formulations where a minimum sensing SNR is imposed to guarantee reliable target estimation~\cite{sivadevuni2023JCAS}, we enforce a sensing-reliability constraint. In the present FBL setting, the blocklength $\eta$ also determines the number of sensing samples collected within one sensing block. Hence, under coherent matched-filter processing, the accumulated sensing SINR is modeled as $\eta\gamma_s$. The resulting problem is
\begin{subequations}\label{eq:opt1} \begin{align} \mathcal{P}_1\!: \min_{\mathbf{W}, \mathbf{v}, \mathbf{r}} & \quad \Psi = \sum_{k \in \mathcal{K}} \Bigg| \frac{r_k}{r^{\text{des}}_{k}} -1\Bigg| \tag{\ref{eq:opt1}} \\ 
\text{s.t.}\quad 
&r_k \!\leq\!B\!\left(\!\log_2(1\!+\!\!\Gamma_k)\!-\!\Omega\!\sqrt{\tfrac{V(\Gamma_k)}{\eta}}\right)\!, \!\!\!\!&&\!\forall k\!\in\!\mathcal{K},\label{eq:opt1JCAS_Opt2} \\ 
&\eta \,\gamma_s \geq \gamma_s^{\mathsf{min}},\!\!&&\!\label{eq:opt1JCAS_Opt3} \\
&\sum_{k \in \mathcal{K}}\!\|\mathbf{w}_k\|_2^2 + \|\mathbf{w}_s\|_2^2 \leq P_{T}, \label{eq:opt1JCAS_Opt4} \\ 
&|v_m|=1,&&\!\forall m\!\!\in\!\!\mathcal{M}, \label{eq:opt1JCAS_Opt5} 
\end{align} 
\end{subequations} 
where $\mathcal{M}:=\{1,\dots,M\}$ denotes the set of RIS reflecting elements and $\mathbf{r}=[r_1,\ldots,r_K]^T$ is the rate vector. Constraint~\eqref{eq:opt1JCAS_Opt2} captures the FBL achievable-rate bound, whose penalty term decreases with increasing $\eta$ and approaches the IBL regime for large blocklengths. Constraint~\eqref{eq:opt1JCAS_Opt3} imposes a minimum accumulated sensing SINR over one sensing block: a smaller instantaneous sensing SINR $\gamma_s$ must be compensated by a larger blocklength, whereas a stronger sensing link allows shorter blocks. This models the latency--sensing tradeoff induced by FBL operation. Finally,~\eqref{eq:opt1JCAS_Opt4} limits the BS transmit power, while~\eqref{eq:opt1JCAS_Opt5} enforces the unit-modulus RIS phase-shift constraint. The QoS requirements in~\eqref{eq:opt1JCAS_Opt2}--\eqref{eq:opt1JCAS_Opt3} are assumed constant within one coherence interval $T_{\mathrm{c}}$.

Problem $\mathcal{P}_1$ is highly nonconvex due to the coupling of $\mathbf{W}$ and $\mathbf{v}$, the unit-modulus constraint, and channel dispersion, and is typically solved with alternating optimization and successive convex approximation (SCA). To address problem $\mathcal{P}_1$, we adopt a solution strategy inspired by~\cite{weinberger2025symbolcountsresilientwireless}. Yet, unlike the approach in~\cite{weinberger2025symbolcountsresilientwireless}, the present formulation explicitly incorporates the sensing SINR constraint, which is inherently coupled with the blocklength parameter $\eta$. To facilitate the application of efficient optimization methods, problem $\mathcal{P}_1$ is reformulated into an equivalent but more tractable representation. This reformulation enables the use of alternating optimization and SCA, which are effective in handling the nonconvex structure induced by FBL constraints. Specifically, we introduce two sets of nonnegative slack variables, $\mathbf{q}=[q_1,\dots,q_K, q_s]$ and $\mathbf{u}=[u_1,\dots,u_K]$, in order to convexify the achievable rate expressions. The resulting optimization problem is expressed as 
\begin{subequations}\label{eq:opt2}
\begin{align} \mathcal{P}_2\!: \min_{\mathbf{W}, \mathbf{v}, \mathbf{r}, \mathbf{q}, \mathbf{u}}
&\quad \Psi = \sum_{k \in \mathcal{K}} \Bigg| \frac{r_k}{r^{\text{des}}_{k}} -1\Bigg| \tag{\ref{eq:opt2}} \\
\text{s.t.}\quad
&r_k \!\leq\! B\!\left( \!\log_2(1\!+\!q_k\!) \!-\! \tfrac{\Omega}{\sqrt{\eta}}\!u_k\! \right), \!\!&&\!\forall k\!\in\!\mathcal{K}, \label{eq:JCAS_Opt2} \\
&\eta \,q_s \geq \gamma_s^\mathsf{min}, \!\!&&\! \label{eq:JCAS_Opt3} \\
&q_k \leq \Gamma_k, \!\!&&\!\forall k\!\in\!\mathcal{K}, \label{eq:JCAS_Optn1}\\
&q_s\leq \gamma_s \!\!&&\!\label{eq:JCAS_Optn2}\\
&u_k \geq \sqrt{V(q_k)}, \!\!&&\!\forall k\!\in\!\mathcal{K},\label{eq:JCAS_Optn3}\\
&\sum_{k \in \mathcal{K}}\|\mathbf{w}_k\|_2^2 + \|\mathbf{w}_s\|_2^2 \leq P_\mathrm{T}, \!\!&&\! \label{eq:JCAS_Opt4} \\ &|v_m|=1, \!\!\!&&\!\forall m\!\in\!\mathcal{M}. \label{eq:JCAS_Opt5}
\end{align} \end{subequations}
It is worth noting that although~\eqref{eq:JCAS_Optn1}~-~\eqref{eq:JCAS_Optn3} remain nonconvex in their original form, they can be convexified via the SCA technique, thereby enabling efficient iterative alternating optimization. The detailed problem reformulation and the alternating optimization algorithm are given in Appendix A.

 %================= Numerical =================% 
\section{Numerical Results} \label{Sim_results}    

We investigate the performance of the proposed ISAC system in a two-dimensional area of $[0,150] \times [-10,40] \,\text{m}^2$. The BS is placed at the origin and operates at a carrier frequency of $\SI{5}{GHz}$ with a bandwidth of $B=\SI{10}{MHz}$. It is equipped with $N_T = 8$ transmit and $N_R = 8$ receive antennas, and its total transmit power is set to $P_\mathrm{T}=\SI{32}{dBm}$.

The scenario involves a moving point target that starts at $[100,-1]$ at $T=0$ and travels upward along the $y$-axis with a constant velocity of $\SI{30}{m/s}$. The target is tracked over a period of $T_\mathsf{coh} = \SI{100}{ms}$, during which the communication user's channels are assumed to remain within one coherence time. The target's channel changes after each transmission and is therefore assumed to have the coherence time $T_\mathsf{sens}$, which depends on the blocklength chosen as defined by (\ref{eq:T_sens}). The radar channel is modeled with a path loss exponent of $\nu_{xy}=4$, reference distance $d_o = \SI{1}{m}$, and RCS of $\SI{1}{m^2}$. The LoS path depends on the target’s instantaneous position and is updated after each transmission to account for target motion. 

Two communication users are located at $[120,4]$ and $[140,0]$. They are served both via direct LoS links and via a RIS positioned at $[135,40]$. The RIS consists of $50 \times 50$ elements with $\lambda_f/4$ spacing, where $\lambda_f$ denotes the wavelength. The LoS communication channels follow a Rician fading model with factor $K_\mathrm{rice}=50$, representing a moderate LoS-dominant propagation environment commonly adopted in RIS-assisted ISAC studies, while the RIS channels are generated according to the correlated model in \cite{weinberger2023ris}. Each user requires a target data rate of $r^{\text{des}}_k = 10 \,\text{Mbps}$ with error probability $\epsilon = 10^{-3}$, under a noise level of $\sigma^2 = \SI{-100}{dBm}$.

During target motion, the sensing beam occasionally crosses the LoS path of each communication user, blocking the user and also creating strong interference. When this occurs, the communication beams are rerouted through the RIS-assisted link, while ensuring that the radar maintains its sensing QoS, defined as a minimum SINR of $\gamma_s^\mathsf{min} = \SI{-5}{dB}$ and $\alpha_v=1000$.

\subsection{Interplay of SI level and blocklength on ISAC Performance}

\begin{figure}[htpb]
  \centering
  \centerline{\includegraphics[width=\linewidth]{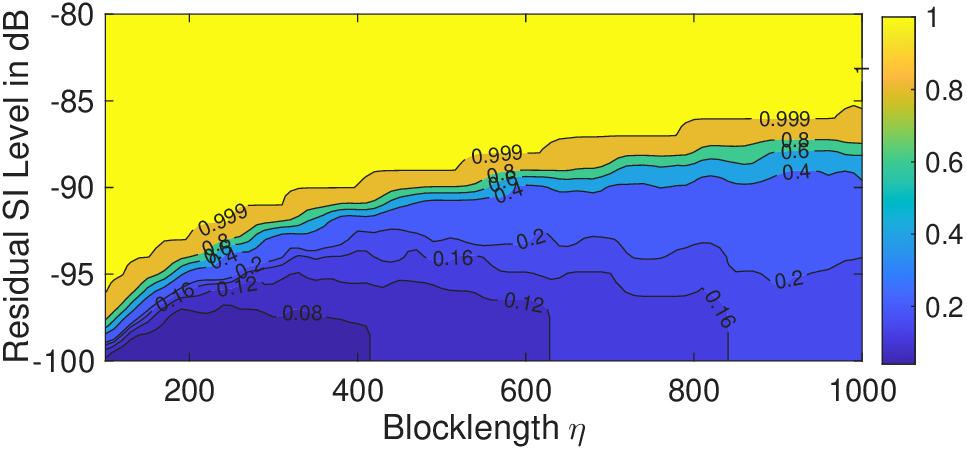}}
\caption{Outage probabilities of the ISAC QoS over the blocklength for different residual SI levels.}\label{fig:OutagevsBL}
  \end{figure}

Fig.~\ref{fig:OutagevsBL} shows how residual SI and FBL jointly drive ISAC QoS outages, with two monotone effects. For fixed blocklength~$\eta$, higher residual SI increases outage by degrading radar SINR~$\gamma_s$, which tightens the sensing constraint~$\eta q_s \ge \gamma_s^{\min}$ and reduces the margin for downlink QoS. For fixed residual SI, increasing~$\eta$ lowers outage because the FBL penalty in~\eqref{eq:JCAS_Opt2} scales as $1/\sqrt{\eta}$ and the sensing constraint in~\eqref{eq:JCAS_Opt3} becomes easier via linear scaling. 
The figure also shows a transition between operating regimes. When residual SI is strong, the system stays outage-limited over essentially all values of~$\eta$, whereas with effective SI cancellation a moderate increase in~$\eta$ is enough to enter a low-outage regime. In other words, increasing~$\eta$ can partially offset residual SI, but only up to the SI floor, which ultimately caps the achievable~$\gamma_s$ and overall reliability. In the low-SI regime, outage can even increase with~$\eta$. Longer blocks imply a longer sensing interval, which raises the likelihood of target-motion mismatch and degrades~$\gamma_s$. That loss can outweigh the reduced FBL penalty, highlighting a trade-off between blocklength efficiency and mobility-induced sensing degradation.

\subsection{ISAC Full-duplex against Half-duplex Performance}
\begin{figure}[t]
  \centering
  \centerline{\includegraphics[width=\linewidth]{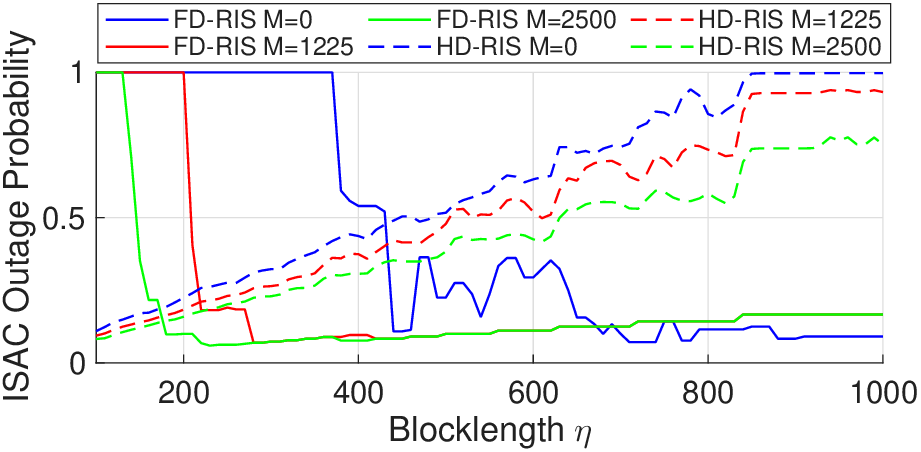}}
\caption{Outage probabilities of the ISAC QoS over the blocklength for no RIS, $35\times35$, and $50\times50$ sized RISs, comparing FD and HD performances for an SI level of \SI{-100}{dB}.}\label{fig:FDvsHD}
  \end{figure}

Fig.~\ref{fig:FDvsHD} shows the ISAC QoS outage probability versus blocklength for FD and HD operation with no RIS and RIS sizes of $35\times35$ and $50\times50$ at an SI level of $\SI{-100}{dB}$. To ensure a fair comparison under FBL constraints, the HD case accounts for the reduced degrees of freedom by using half the bandwidth, $B_{\mathsf{HD}}=B/2$, while excluding SI from the sensing SINR. Consequently, HD operation halves both the effective communication rate and the sensing duration, resulting in fewer sensing samples.

At short blocklengths, FD operation exhibits high outage due to the severe FBL penalty combined with residual SI. Once the blocklength exceeds a threshold, this penalty becomes negligible and the outage probability drops sharply, indicating a transition to a SINR-limited regime. RIS deployment shifts this transition to smaller blocklengths by improving the effective channel conditions; notably, a $35\times35$ RIS approximately halves the blocklength required to meet the ISAC QoS targets, directly reducing the required sensing duration.

In contrast, HD outage increases almost linearly with blocklength, reflecting its dominant limitation from time-sharing rather than SI. RIS assistance provides only marginal gains in HD, as the effective-rate reduction and shortened sensing duration fundamentally limit reliability. Beyond the FD transition point, FD consistently outperforms all HD benchmarks.

For large blocklengths, HD outage eventually increases due to longer sensing intervals, during which target motion degrades sensing performance. This effect is most pronounced without RIS support, where the absence of RIS phase optimization allows more frequent sensing updates. As RIS optimization primarily benefits communication, it can delay sensing reconfiguration and, in the long-blocklength regime, lead to inferior ISAC reliability compared to the no-RIS case. Overall, the results reveal a regime-dependent trade-off: HD is preferable at very short blocklengths, whereas FD becomes superior once the finite-blocklength penalty is sufficiently small, with the transition governed by the SI level and the imposed ISAC QoS requirements.

% \begin{figure}[t]
%   \centering
%   \centerline{\includegraphics[width=\linewidth]{Figures/RadarEcho.eps}}
% \caption{Radar outage probability and variance of the radar echo SINR over the  blocklength $\eta$ for different levels of SI cancellation.}\label{fig:RadarEcho}
%   \end{figure}
% \begin{figure}[t]
%     \centering
%     \begin{subfigure}[b]{0.2135\textwidth}
%         \centering
%         \includegraphics[width=\textwidth]{Figures/SqSens-120.pdf}
%         \caption{$\rho=-120$~dB}
%         \label{fig:SqSens120}
%     \end{subfigure}
%     \hfill
%     \begin{subfigure}[b]{0.24\textwidth}
%         \centering
%         \includegraphics[width=\textwidth]{Figures/SqSens-118.pdf}
%         \caption{$\rho=-118$~dB}
%         \label{fig:SqSens118}
%     \end{subfigure}
%     \caption{Radar sensing performance: comparison of outage probability and radar echo SINR variance for two levels of SI cancellation. Left: ideal SI ($\rho=-120$~dB), right: higher residual SI ($\rho=-118$~dB).}
%     \label{fig:SqSensComparison}
% \end{figure}

 %================= Conclusions =================% 
\section{Conclusion} \label{Conclusion}
    This work shows that RIS-assisted FD ISAC can balance communication and sensing performance under FBL and residual SI. Joint optimization of ISAC BS beamforming and RIS beamsteering reduces service adaptation gaps while preserving reliable radar detection of a moving target. Numerical results further indicate a regime-dependent design rule: blocklength is not only a reliability parameter but also a control variable that governs the tradeoff among communication efficiency, sensing freshness, and SI. Longer blocks can yield substantial gains with effective SI suppression in FD operation, yet may degrade sensing when update latency dominates, producing non-monotonic reliability. Overall, optimal ISAC design requires joint consideration of blocklength, duplexing mode, and sensing dynamics, rather than optimizing any single factor in isolation.
\appendix
\subsection{SCA-Based Alternating Optimization}
In alternating optimization, the phase vector $\mathbf{v}$ is first fixed while updating beamforming directions. To convexify problem $\mathcal{P}_2$, the constraints in~\eqref{eq:JCAS_Optn1} -~\eqref{eq:JCAS_Optn3} are linearized using a first-order Taylor expansion at $(\tilde{\mathbf{w}},\tilde{\mathbf{q}})$, following~\cite{weinberger2025symbolcountsresilientwireless,weinberger2023ris}. The convex surrogate of~\eqref{eq:JCAS_Optn1} is
\begin{align}\label{eq:SINR_convex1}
	\sum_{k' \in \mathcal{K}\setminus\{k\}} |\widehat{\mathbf{h}}_{k}^H \mathbf{w}_{k'}|^2 +
    |\widehat{\mathbf{h}}_{k}^H \mathbf{w}_s|^2  
    + \sigma^2_k
    + \frac{|\widehat{\mathbf{h}}_k^H \tilde{\mathbf{w}}_k|^2}{\tilde{q}_k^2}q_k \nonumber \\ 
    - \frac{2\, \Re \{ \tilde{\mathbf{w}}_k^H \widehat{\mathbf{h}}_{k}\widehat{\mathbf{h}}_{k}^H \mathbf{w}_k \}}{\tilde{q}_k} 
    \leq 0, \quad \forall k \in \mathcal{K}.
\end{align}
Similarly,~\eqref{eq:JCAS_Optn1} yields
\begin{equation}\label{eq:SINR_convex2}
\rho \widetilde{P}_\mathrm{T}(\mathbf{W}) + \sigma_s^2 
+ \frac{\|\mathbf{G}\tilde{\mathbf{w}}_s\|_2^2}{\tilde{q}_s^2}q_s
- \frac{2\, \Re\{\tilde{\mathbf{w}}_s^H \mathbf{G}^H \mathbf{G}\tilde{\mathbf{w}}_s\}}{\tilde{q}_s} 
\leq 0.
\end{equation}
For the dispersion term $\sqrt{V(q_k)}$ in~\eqref{eq:JCAS_Optn3}, first-order expansion at $\tilde{q}_k$ gives~\cite{xu2025maxmin}
\begin{align}\label{eq:ChanDispApp}
\sqrt{V(q_k)} \leq
\sqrt{1 - (1+\tilde{q}_k)^{-2}} 
+ (1+\tilde{q}_k)^{-3} \nonumber\\ 
\times (1 - (1+(\tilde{q}_k)^{-2}))^{-\frac{1}{2}} (q_k - \tilde{q}_k) 
\triangleq U_k(q_k).
\end{align}

Thus, $\mathcal{P}_2$ is approximated by the convex problem
\begin{subequations}\label{eq:opt3}
\begin{align}
\mathcal{P}_3\!: 
\min_{\mathbf{W}, \mathbf{v}, \mathbf{r}, \mathbf{q},  \mathbf{u}} & \quad 
    \Psi  \tag{\ref{eq:opt3}} \\
\text{s.t.}\quad 
&\eqref{eq:JCAS_Opt2},~\eqref{eq:JCAS_Opt3},~\eqref{eq:JCAS_Opt4},~\eqref{eq:JCAS_Opt5},\\
&\eqref{eq:SINR_convex1},~\eqref{eq:SINR_convex2}, \nonumber \\
&u_k \geq U_k(q_k).\label{eq:opt3_eq1}
\end{align}
\end{subequations}
Problem $\mathcal{P}_3$ is convex and solved iteratively via SCA. At iteration $z$, beamforming variables are collected in 
$\mathbf{\Lambda}_z^w = [\mathbf{w}_z^T , \mathbf{\kappa}_z^T]^T$, with $\mathbf{\kappa}_z = [\mathbf{r}_z^T, \mathbf{q}_z^T, \mathbf{u}_z]^T$. The solution $\hat{\mathbf{\Lambda}}_z^w$ is obtained using local approximations at point $\tilde{\mathbf{\Lambda}}_z^w$.

For the RIS phase-shift optimization, beamforming vectors are fixed. Again,~\eqref{eq:JCAS_Optn1} is convexified, but this time at $(\tilde{\mathbf{v}},\tilde{\mathbf{q}})$. 
% With
% \begin{equation}
% \widehat{\mathbf{h}}_{k'}^H \mathbf{w}_k 
% = \mathbf{w}_k^H \mathbf{h}_{k'} + \mathbf{w}_k^H \mathbf{G}_{k'} \mathbf{v} 
% = \beta_{k',k},
% \end{equation}
The communication SINR constraints are linearized as~\cite{weinberger2025symbolcountsresilientwireless,weinberger2023ris}
\begin{align}\label{apprxV}
&\sum_{k'\in\mathcal{K}\setminus\{k\}}|\widehat{\mathbf{h}}_{k}^H \mathbf{w}_{k'} |^2 + |\widehat{\mathbf{h}}_{k}^H \mathbf{w}_s|^2  + \sigma_k^2 
-\frac{|{\widehat{\mathbf{h}}_{k}^H \mathbf{w}_k }|^2}{\tilde{q}_k } \nonumber\\
&-\frac{2}{\tilde{q}_k} \Re\!\big\{ (\mathbf{w}_k^H \mathbf{h}_k + \mathbf{w}_k^H \mathbf{H}^H \text{diag}(\mathbf{g}_k) \tilde{\mathbf{v}})^*  \mathbf{w}_k^H \mathbf{H}^H \text{diag}{(\mathbf{g}_k)} \nonumber\\&(\mathbf{v} - \tilde{\mathbf{v}}) \big\}+\frac{|\widehat{\mathbf{h}}_{k}^H \mathbf{w}_k|^2}{\tilde{q}_k^2} ( q_k - \tilde{q}_k ) \leq 0,  \quad \forall k \in \mathcal{K}.
\end{align}
The unit-modulus constraint in~\eqref{eq:JCAS_Opt5} is addressed by the penalty method~\cite{weinberger2023ris}, adding
\begin{equation}
\Phi = \alpha_{v}\sum_{m=1}^{M}\Re\!\left\{2\tilde{v}_m^*v_m - |\tilde{v}_m|^2\right\},
\end{equation}
to the objective, with $\alpha_v \gg 1$.

The RIS optimization becomes
\begin{subequations}\label{eq:opt4}
\begin{align}
\mathcal{P}_4\!: 
	\underset{\mathbf{v},\mathbf{r},\mathbf{q},\mathbf{u}}{\min}  \quad & \Psi - \Phi \tag{\ref{eq:opt4}} \\
	\text{s.t.} \quad
        &\eqref{eq:JCAS_Opt2},~\eqref{eq:JCAS_Opt3},~\eqref{eq:JCAS_Opt4},~\eqref{eq:JCAS_Opt5},\\
        &\eqref{eq:SINR_convex1},~\eqref{eq:SINR_convex2},~\eqref{eq:opt3_eq1}.
\end{align}
\end{subequations}
Since $\mathcal{P}_4$ has the same SCA structure as $\mathcal{P}_3$, it is solved iteratively with $\mathbf{\Lambda}_z^v = [\mathbf{v}_z^T, \mathbf{\kappa}_z^T]^T$.

The detailed steps of the alternating optimization
are illustrated in Algorithm 1, which alternatingly optimizes ISAC BS beamformers and RIS beamsteering.
\begin{algorithm}[H]
\setlength{\topsep}{0pt}
\setlength{\parskip}{0pt}
\footnotesize
\caption{Alternating Optimization}\label{alg}
\begin{tikzpicture}[>={Latex[length=5pt]}, baseline=(current bounding box.north)]\label{alg:AltOpt}
	\footnotesize
	\tikzset{myRect/.style={draw,rectangle,minimum width=0.5cm, minimum height=0.25cm,align=center}}
	\tikzset{myDiam/.style={draw,diamond,aspect=1.5,minimum width=1.25cm, minimum height=0.9cm,align=center}}
	\tikzset{myArrow/.style={->,draw,line width=0.25m}}
	\tikzset{myDot/.style={draw,minimum size=0.1,circle,fill,scale=0.15}}
	
	\node[align=center](input) at (-1,-1.85){ Input:\\ $\tilde{\mathbf{w}},\tilde{\mathbf{v}},\tilde{\mathbf{\kappa}},$\\ $T_\mathsf{c}$,${\alpha_v}$};
	
	\node[myRect](initLam) at ($(input)+(0,2.45)$){Create\\$\tilde{\mathbf{\Lambda}}_z^o$};
	%\node (initLamTxt) at (initLam){};

	\node[myDiam] (o1) at (0,0){};
	\node[] (o1txt) at (o1){$o=w$};
	\node[myRect] (P2) at ($(o1)+(2.625,0.3)$){$\hat{\mathbf{\Lambda}}_{z+1}^w \leftarrow$ solve $\mathcal{P}_3$};
	\node[myRect] (P3) at ($(o1)+(2.625,-0.3)$){$\hat{\mathbf{\Lambda}}_{z+1}^v \leftarrow$ solve $\mathcal{P}_4$};
	\node[myDiam] (oT) at (6.,-0.75){};
	\node[] (oTText) at (oT){$T{<}\:T_\mathsf{c}$};
	\node[myDiam] (o2) at (4.555,-1.5){};
	\node[] (o2txt) at (o2){$o=w$};
	\node[myRect,align=center] (Lambw) at ($(o2)+(-2.45,0.45)$){$[\tilde{\mathbf{w}}^T, \tilde{\mathbf{\kappa}}^T]^T \leftarrow \hat{\mathbf{\Lambda}}_z^w$ ,\\$\tilde{\mathbf{\Lambda}}_z^v \leftarrow [\tilde{\mathbf{v}}^T, \tilde{\mathbf{\kappa}}^T]^T $};
	\node[myRect,align=center] (Lambv) at ($(o2)+(-2.45,-0.45)$){$[\tilde{\mathbf{v}}^T, \tilde{\mathbf{\kappa}}^T]^T \leftarrow \hat{\mathbf{\Lambda}}_z^v$ \\ $\tilde{\mathbf{\Lambda}}_z^w \leftarrow [\tilde{\mathbf{w}}^T, \tilde{\mathbf{\kappa}}^T]^T$};
	
%	\node[myDiam](Psi) at ($(oT)+(-0.45,-1.025)$){};
%	\node(Psitxt) at (Psi) {$\Uppsi<\tau$};

	\draw[->] (input.north) |- ($(input.north)!0.25!(initLam.240)$) -|node[pos=0.76,anchor=210]{$\,\,\,\phantom{.}_{o\leftarrow w}$}node[pos=0.535,anchor=210]{$\phantom{.}_{T{\leftarrow 0}}$}node[pos=0.655,anchor=210]{$\phantom{.}_{z{\leftarrow 0}}$} (initLam.240);
	\draw[->] (initLam.south) |- (o1.west);

	\draw[->] (o1.east) --node[myDot,pos=1]{} ($(o1.east)+(0.1,0)$) |-node[pos=0.65,above]{\tiny$\mathsf{T}$} (P2.west) ;
	\draw[->] (o1.east) -- ($(o1.east)+(0.1,0)$)  |-node[pos=0.65,above]{\tiny$\mathsf{F}$} (P3.west) ;
	
	\node[myDot] (incrementDot) at ($(oT.north)-(1.9,-0.25)$){};
	\node[myDot] (incrementDot2) at ($(oT.north)-(1.75,-0.25)$){};
    \draw     (incrementDot) -- (incrementDot2);

    \draw[->] (P2.east) -| (incrementDot);
	\draw[->] (P3.east) -| (incrementDot);
	
	\draw[->] (incrementDot2) |-node[pos=0.775,below]{$\phantom{.}_{z\,{\leftarrow}{z+1}}$}node[pos=0.775,above]{ \tiny${T{\leftarrow}{T\hspace{-0.075cm}+\hspace{-0.075cm}T_\mathsf{sens}}}$} (oT.west);
\draw[->] (incrementDot2) |-node[pos=1,anchor=180]{Output:\\$\hat{\mathbf{w}}_z,\hat{\mathbf{v}}_z$}($(incrementDot2)+(0.5,0.5)$) ;
	
	%\draw[->] (oT.north) |- node[pos=0.1,right]{\tiny$\mathsf{T}$} node[pos=0.65,below]{$\tilde{\mathbf{\Lambda}}_z^o \leftarrow \hat{\mathbf{\Lambda}}_z^o$}  ($($(oT.north)!0.5!(o1.north)$) + (0,0.4) $) -| (o1.north);
	
	\draw[->] (oT.south) |- node[pos=0.35,anchor=-35]{\tiny$\mathsf{F}$}  (o2.east); %($(oT.west)!0.51!(o2.north)$)
	
	\draw[->] (o2.west) --node[myDot,pos=1]{} ($(o2.west)-(0.125,0)$) |-node[pos=0.65,above]{\tiny$\mathsf{T}$} (Lambw.east) ;
	\draw[->] (o2.west) -- ($(o2.west)-(0.125,0)$)  |-node[pos=0.65,above]{\tiny$\mathsf{F}$} (Lambv.east) ;
	
	\node[myDot] (reassignDot) at  ($(o2)-(4.39,0)$) {};
	\draw[->] (Lambw.west) -|node[pos=0.25,above]{$\phantom{.}_{o\leftarrow v}$} (reassignDot) ;
    \draw[->] (Lambv.west) -|node[pos=0.25,above]{$\phantom{.}_{o\leftarrow w}$} (reassignDot) ;

    \draw[->] (reassignDot) -|node[pos=0.65,anchor=-25]{} (o1.south);
	
	\path (oT.east) |-node[pos=0.85,anchor=125,align=center](a){} node[pos=0.15,anchor=35,align=center](b){\tiny$\mathsf{T}$} ($(oT.east)-(0.0,1)$);% -- (Psi.east); %(oT.east)!0.5!
	
\draw[->] (oT.east) -- ($(oT.east)-(0.0,1)$);
	\node[align=center] at ($(a)+(-0.1,-0.15)$){Stop};
	
%Output:\\$\hat{\mathbf{w}}_z,\hat{\mathbf{v}}_z$
%	\draw[->] (Psi.south) --node[pos=0.25,right]{\tiny$\mathsf{T}$} ($(Psi.south)+(0,-0.4)$);
	
%	\node at ($(Psi.south)+(0,-0.65)$) {Stop};
	
\end{tikzpicture}
\end{algorithm}

 %================= References =================%
\bibliographystyle{IEEEtran}
\bibliography{bibliography}

@misc{li2025lightweightframeworkintegratedsensing,
      title={A Lightweight Framework for Integrated Sensing and Communications with {RIS}}, 
      author={Chu Li and Kevin Weinberger and Aydin Sezgin},
      year={2025},
      eprint={2511.04448},
      archivePrefix={arXiv},
      primaryClass={eess.SP},
      url={https://arxiv.org/abs/2511.04448}, 
}

@article{liu2020joint,
  title={Joint transmit beamforming for multiuser {MIMO} communications and {MIMO} radar},
  author={Liu, Xiang and Huang, Tianyao and Shlezinger, Nir and Liu, Yimin and Zhou, Jie and Eldar, Yonina C},
  journal={IEEE Trans. Signal Process.},
  volume={68},
  pages={3929--3944},
  year={2020},
  publisher={IEEE}
}

@ARTICLE{jing2024ISAC,
  author={Jing, Xiaoye and Liu, Fan and Masouros, Christos and Zeng, Yong},
  journal={IEEE Trans. Wirel. Commun.}, 
  title={ISAC From the Sky: {UAV} Trajectory Design for Joint Communication and Target Localization}, 
  year={2024},
  volume={23},
  number={10},
  pages={12857-12872},
  doi={10.1109/TWC.2024.3396571}}

@ARTICLE{reifert2022comeback,
  author={Reifert, Robert-Jeron and Roth, Stefan and Ahmad, Alaa Alameer and Sezgin, Aydin},
  journal={IEEE Trans. Veh. Technol.}, 
  title={Comeback Kid: Resilience for Mixed-Critical Wireless Network Resource Management}, 
  year={2023},
  volume={72},
  number={12},
  pages={16177-16194},
  doi={10.1109/TVT.2023.3296977}}

@article{basar2019wireless,
  title={Wireless communications through reconfigurable intelligent surfaces},
  author={Basar, Ertugrul and Di Renzo, Marco and De Rosny, Julien and Debbah, Merouane and Alouini, Mohamed-Slim and Zhang, Rui},
  journal={IEEE Access},
  volume={7},
  pages={116753--116773},
  year={2019},
  publisher={IEEE}
}

@inproceedings{weinberger2023ris,
  title={{RIS}-enhanced Resilience in Cell-Free {MIMO}},
  author={Weinberger, Kevin and Reifert, Robert-Jeron and Sezgin, Aydin and Basar, Ertugrul},
  booktitle={WSA \& SCC},
  pages={},
  year={2023},
  organization={VDE}
}

@ARTICLE{polyanskiy2010FBL,
  author={Polyanskiy, Yury and Poor, H. Vincent and Verdu, Sergio},
  journal={IEEE Trans. Inf. Theory}, 
  title={Channel Coding Rate in the Finite Blocklength Regime}, 
  year={2010},
  volume={56},
  number={5},
  pages={2307-2359},
  doi={10.1109/TIT.2010.2043769}}

@misc{weinberger2025symbolcountsresilientwireless,
      title={When Every Symbol Counts: Resilient Wireless Systems Under Finite Blocklength Constraints}, 
      author={Kevin Weinberger and Aydin Sezgin},
      year={2025},
      eprint={2506.21664},
      archivePrefix={arXiv},
      primaryClass={eess.SP},
      url={https://arxiv.org/abs/2506.21664}, 
}

@ARTICLE{xu2025maxmin,
  author={Xu, Jiawei and Clerckx, Bruno},
  journal={IEEE Trans. Commun.}, 
  title={Max-Min Fairness and PHY-Layer Design of Uplink {MIMO} Rate-Splitting Multiple Access With Finite Blocklength}, 
  year={2025},
  volume={73},
  number={5},
  pages={3671-3682},
  doi={10.1109/TCOMM.2024.3490492}}

@ARTICLE{liu2022ISAC,
  author={Liu, An and Huang, Zhe and Li, Min and Wan, Yubo and Li, Wenrui and Han, Tony Xiao and Liu, Chenchen and Du, Rui and Tan, Danny Kai Pin and Lu, Jianmin and Shen, Yuan and Colone, Fabiola and Chetty, Kevin},
  journal={IEEE Commun. Surv. Tutor.}, 
  title={A Survey on Fundamental Limits of Integrated Sensing and Communication}, 
  year={2022},
  volume={24},
  number={2},
  pages={994-1034},
  doi={10.1109/COMST.2022.3149272}}

@ARTICLE{zhang20196G,
  author={Zhang, Zhengquan and Xiao, Yue and Ma, Zheng and Xiao, Ming and Ding, Zhiguo and Lei, Xianfu and Karagiannidis, George K. and Fan, Pingzhi},
  journal={IEEE Veh. Tech. Mag.}, 
  title={{6G} Wireless Networks: Vision, Requirements, Architecture, and Key Technologies}, 
  year={2019},
  volume={14},
  number={3},
  pages={28-41},
  doi={10.1109/MVT.2019.2921208}}

@INPROCEEDINGS{xiao2021FD,
  author={Xiao, Zhiqiang and Zeng, Yong},
  booktitle={13th WCSP}, 
  title={Full-Duplex Integrated Sensing and Communication: Waveform Design and Performance Analysis}, 
  year={2021},
  volume={},
  number={},
  pages={1-5},
  doi={10.1109/WCSP52459.2021.9613663}}

@ARTICLE{xiao2022FD,
  author={Xiao, Zhiqiang and Zeng, Yong},
  journal={IEEE J. Sel. Areas Commun.}, 
  title={Waveform Design and Performance Analysis for Full-Duplex Integrated Sensing and Communication}, 
  year={2022},
  volume={40},
  number={6},
  pages={1823-1837},
  doi={10.1109/JSAC.2022.3155509}}

@Article{nguyen2021SI,
AUTHOR = {Nguyen, Ba Cao and The Dung, Le and Nguyen, Huu Minh and Kim, Taejoon and Kim, Young-Il},
TITLE = {Impacts of Residual Self-Interference, Hardware Impairment and Cascade {Rayleigh} Fading on the Performance of Full-Duplex Vehicle-to-Vehicle Relay Systems},
JOURNAL = {Sensors},
VOLUME = {21},
YEAR = {2021},
NUMBER = {16},
ARTICLE-NUMBER = {5628},
URL = {https://www.mdpi.com/1424-8220/21/16/5628},
PubMedID = {34451069},
ISSN = {1424-8220},
DOI = {10.3390/s21165628}
}

@INPROCEEDINGS{sivadevuni2023JCAS,
  author={Sivadevuni, Srivardhan and Lotfi, Fatemeh and Ahmad, Bilal and Weinberger, Kevin and Sezgin, Aydin},
  booktitle={IEEE 9th CAMSAP}, 
  title={Preparing for the Inevitable: Preventing Outages Using Resilient {RIS}-Assisted {JCAS}}, 
  year={2023},
  volume={},
  number={},
  pages={241-245},
  doi={10.1109/CAMSAP58249.2023.10403515}}

@ARTICLE{durisi2016ShortPack,
  author={Durisi, Giuseppe and Koch, Tobias and Popovski, Petar},
  journal={Proc. IEEE}, 
  title={Toward Massive, Ultrareliable, and Low-Latency Wireless Communication With Short Packets}, 
  year={2016},
  volume={104},
  number={9},
  pages={1711-1726},
  doi={10.1109/JPROC.2016.2537298}}

@INPROCEEDINGS{umra2025RL,
  author={Umra, Adam and Ahmed, Aya Mostafa and Sezgin, Aydin},
  booktitle={Proc. EuCNC/6G Summit}, 
  title={Towards Smarter Sensing: {2D} Clutter Mitigation in {RL}-Driven Cognitive {MIMO} Radar}, 
  year={2025},
  volume={},
  number={},
  pages={1-6},
  doi={10.1109/EuCNC/6GSummit63408.2025.11036963}}

@ARTICLE{trevlakis2023localization,
  author={Trevlakis, Stylianos E. and Boulogeorgos, Alexandros-Apostolos A. and Pliatsios, Dimitrios and Querol, Jorge and Ntontin, Konstantinos and Sarigiannidis, Panagiotis and Chatzinotas, Symeon and Di Renzo, Marco},
  journal={IEEE Open J. Commun. Soc.}, 
  title={Localization as a Key Enabler of {6G} Wireless Systems: A Comprehensive Survey and an Outlook}, 
  year={2023},
  volume={4},
  number={},
  pages={2733-2801},
  doi={10.1109/OJCOMS.2023.3324952}}
	
\end{document}